\documentclass[aps,pre,twocolumn,floats,floatfix,english]{revtex4}
\usepackage[T1]{fontenc}
\usepackage[latin1]{inputenc}
\usepackage{amsmath}
\usepackage{graphicx}
\usepackage{setspace}
\usepackage{epsfig}
\makeatletter

\providecommand{\LyX}{L\kern-.1667em\lower.25em\hbox{Y}\kern-.125emX\@}

\usepackage{babel}
\makeatother
\begin{document}
\title{Searching good strategies in adaptive minority games}

\author{Marko Sysi-Aho}
\author{Anirban Chakraborti}
\author{Kimmo Kaski}

\affiliation{Laboratory of Computational Engineering, Helsinki
University of Technology, \\
P. O. Box 9203, FIN-02015 HUT, Finland.}

\begin{abstract}
In this paper we introduce adaptation mechanism based on genetic
algorithms in minority games. If agents find their performances too 
low, they modify their strategies in hope to improve their performances 
and become more successful. One aim of this study is to find out what 
happens at the system as well as at the individual agent level. 
We observe that adaptation remarkably tightens the competition among 
the agents, and tries to pull the collective system into a state 
where the aggregate utility is the largest. We first make a brief
comparative study of the different adaptation mechanisms and then
present in more detail parametric studies. These different adaptation
mechanisms broaden the scope of the applications of minority games to
the study of complex systems.
 \end{abstract}

\maketitle
\section{Introduction}
Various systems of natural and societal origin show complex behaviour, 
which can be attributed to competition among interacting agents for 
scarce resources and their adaptation to continuously changing 
environment \cite{parisi,huberman,nowak,lux,arthur}. Such agents 
could be diverse in form, function and capability, for example, 
cells in an immune system or firms in a financial market, so that
in the studies one should first focus on the capabilities of 
individual agents to understand better the nature of interactions 
between large number of agents. The behaviour of an agent may 
be considered as a collection of rules governing {\it responses} 
to {\it stimuli}. In order to model these complex adaptive system, 
a major concern is the selection and representation of the stimuli 
and responses, through which the behaviour and strategies of the 
agents are determined. In a model, the rules of action serve 
as a direct way to describe the strategies of agents, and one studies 
their behaviour by monitoring the effect of rules acting sequentially. 
As mentioned above there is another key process to be included to 
the model, namely {\it adaptation}, which in biology serves as a 
mechanism by which an organism tries to make itself fit to changing 
environment but the timescale over which the agents adapt vary 
from system to system.

What makes these systems fascinatingly complex, is the fact that the 
environment of a particular agent includes other adaptive agents,
all of them competing with each others. Thus, a considerable amount 
of an agent's effort goes in adaptation and reaction to the other 
agents. This feature is the main source of interesting temporal 
patterns and emergent behaviour these systems produce. In fact this 
kind of adaptive systems are far more complicated than systems in which
agents just react following some fixed rules of strategy and foresight for
outcome as a consequence of their behaviour. This latter case can be
tackled with the traditional game theory \cite{game} since it studies 
consistent patterns in behavioural equilibrium that induce no further 
interactions. In the adaptive case, however, further interactions emerge
during the evolution of the system, thus rendering the applicability 
of the traditional game theory difficult if not impossible. 

In this paper, we will study a simple game model, in which 
agents adapt dynamically to be competitive and perform better.  
In such a model the strategies, which an agent uses to decide 
the course of action, must be very good or best for the agents to 
survive -- similar to the idea of ``survival of the fittest'' in biology. 
So just like an organism adapts itself to its natural environment, 
we propose that the agents of the game adapt themselves by modifying 
their strategies from time to time, depending on their current 
performances. For this purpose we borrow the concept of genetic 
crossover from biology and use it to modify the strategies of agents 
in the course of the game, in the same way as in genetic algorithms 
\cite{holland,goldberg,lawrence}. More specifically we apply this 
adaptation scheme to the minority game, introduced by Challet and 
Zhang \cite{challet1,challet2,riolo,cavagna,lamper}. Although the 
behaviour of this minority game is believed to expose a number of 
important characteristics of complex evolving systems, one of its 
weaknesses is that agents have limited possibilities in improving 
to their own performance whereas in real competitive environment 
attempts to improve ones skills continuously are imperative. 
Our adaptation scheme \cite{Marko1,Marko2} proposes a natural 
and simple way to take this essential feature into account, and its 
application turns out to yield results quite different from 
earlier studies of the basic minority game and its variants 
\cite{challet1,challet3,li1,li2}.

This paper is organized such that in the next section we briefly 
introduce our minority game model together with various 
adaptation mechanisms or strategy changes based on one-point 
genetic crossover. This is followed by the results section, 
where we first compare comprehensive computer simulation results of 
these strategies and then focus on analysing results of the model of 
one-point genetic crossover with offsprings replacing parents, 
and of hybridized one-point genetic crossover in which the two new 
strategies replace the two worst strategies. Finally we draw 
conclusions.      

\section{Model}
Let us start by briefly describing the basic minority game (BG) model of
Challet and Zhang \cite{challet1,challet2}. It consists of an odd number 
of agents $N$ who can perform only two actions denoted here by $0$ 
or $1$, at a given time $t$. An agent wins a round of game if it 
is one of the members of the minority group. All the agents are 
assumed to have access to finite amount of ``global'' information, in 
the form of a common bit-string {}``memory'' of the $M$ most recent 
outcomes of the game. With this there are  $2^M$ possible 
``history'' bit-strings. Now, a {}``strategy'' consists of two 
possible responses, which in the binary sense are an action $0$ 
or the opposite action $1$ to each possible history bit-strings. 
Thus, there are $2^{2^{M}}$ possible strategies constituting the 
whole {}``strategy space'' $\Omega$, from which each agent 
picks $S$ strategies at random to form its own pool $\Omega_i$, 
where $i=1,...,N$ denotes an agent number. Each time the game is 
played, time $t$ is incremented by unity and one {}``virtual'' point 
is assigned to the strategies that have predicted the correct outcome 
and the best strategy is the one which has the highest virtual point 
score. The performance of an agent is measured by the number of times 
the agent wins, and the strategy, which the agent used to win, 
gets a {}``real'' point. The number of agents which choose one 
particular action, changes with time and is denoted by $x_t$.

In order to describe the collective behaviour of the agents, we 
define the concept of the scaled utility as a function of $x_t$, 
in the following way:

\begin{equation}
U(x_t)=[(1-\theta(x_t-x_M))x_t+\theta(x_t-x_M)(N-x_t)]/{x_M},
\label{eq1}
\end{equation}

\noindent where $x_M=(N-1)/2$ is the maximum number of agents who 
can win, and

\begin{displaymath}
\theta(x_t-x_M)=\left\{ \begin{array}{ll}
              0 & \textrm{ when $x_t \le x_M$} \\
             1 & \textrm{ when $x_t > x_M$},
\end{array}\right.
\end{displaymath}
 
\noindent is the Heaviside's unit step function. 
When $x_t= x_M$ or $x_t=x_M+1$, the scaled utility of the system
is maximum $U_{max}=1$, as the highest number of agents win. 
The system is more efficient when the deviations from the maximum 
total utility $U_{max}$ are small, or in other words, the 
fluctuations in $x_t$ around the mean ($N/2$) become small.

At the level of individual agents, their performances in the basic
minority game evolve such that the agents who begin to perform badly 
do not improve as the time evolves and those who do well, continue 
doing so \cite{challet1,Marko1}. This indicates that by chance well
performing agents were blessed with good strategies while badly
performing agents got bad strategies. Although this is what may
happen in some real environments, there are other competitive
environments, in which individual agents try to adapt themselves
to do better or to survive. However, being good at one moment does 
not guarantee that one would stay good later. In fact, there are 
many examples in business, sports, etc. which show that those 
who have decided to rest on their laurels have been superseded 
by those who have decided to adapt and fight back, and do so 
persistently. This feature of dynamic competition needs to be 
included in the model, and it can be simply realized by allowing 
agents to modify strategies in their individual pools.  
How well an agent does then in reality, depends on the agent's 
capabilities and skills, and how the agent refines its strategies.  
For the adaptation or strategy modification scheme we have chosen  
genetic algorithms \cite{lawrence}, which have turned out to be
useful in various optimisation problems. Within the framework of 
the minority game this adaptation scheme is realised by letting 
agents check their performances at time interval $\tau$,
and if an agent finds that it is among the worst perfoming fraction
$n$ (where $0<n<1$), it modifies its strategies by applying genetic 
operands to its strategy pool \cite{Marko1,Marko2}. Here the quantity 
$\tau$ describes a time scale that characterizes the adaptation rate
of agents in the system. Hence it can vary on a wide range for 
systems of natural origin to systems of societal nature.  

In the genetic adaptation schemes we have used in this study, an 
agent chooses two ``parents'' from its current pool of strategies 
$\Omega_i(t) \subset \Omega$, and draws a random number 
(uniformly distributed) to determine the crossover point. 
Then the parts of the strategies, above and below this point are 
interchanged to produce two new strategies called the ``offsprings''. 
In addition to this, there are various choises as for which 
strategies are selected as the parents and also which strategies are 
replaced by the reproduced offsprings. The mechanism which works 
the best depends on the circumstances and can vary from system 
to system. In some cases it is possible that saving the parent 
strategies would threaten the success of the newborn strategies 
or create too stiff competition amongst the strategies leading to 
possible disorder, and in other cases the opposite might happen. 
In this study, we have considered four different adaptation schemes 
of first selecting from the strategy pool of an agent, the parent 
strategies to perform genetic crossover for producing offsprings, and 
then selecting the two old strategies that are to be substituted by 
the offspring strategies: 
\begin{description}
\item{(a)} Two parent strategies from the agent's strategy pool are 
drawn at {\it random} and after crossover these parents are substituted
with the two new strategies (offsprings). This procedure is called One 
Point Genetic (OPG) crossover mechanism with parents killed.

\item{(b)} Two parent strategies from the agent's strategy pool are 
drawn at {\it random} and after crossover the two worst performing 
strategies in the strategy pool are substituted with the two new 
strategies (offsprings) while the parent strategies are saved. This 
procedure is called One Point Genetic crossover mechanism with parents 
saved.

\item{(c)} Two {\it best} strategies are chosen from the agent's 
strategy pool as parents and after crossover these parents are 
substituted with the two new strategies (offsprings). This procedure 
is called the Hybridized Genetic (HG) crossover mechanism with 
parents killed.

\item{(d)} Two {\it best} strategies are drawn from the agent's 
strategy pool as parents and after crossover the two worst performing 
strategies in the strategy pool are substituted with the two new 
strategies (offsprings) while the parent strategies are saved. 
This procedure is called the Hybridized Genetic crossover 
mechanism with parents saved.
\end{description}

It should be noted, however, that there are a number of other 
adaptation possibilities, but the ones presented here could be 
considered in a loose sense to bear some resemblance with reality. 
From the point of view of choosing parents, schemes (a) and 
(b) -- being random -- correspond to ``democratic'' or equal opportunity 
reproduction, while schemes (c) and (d) are ``elitist'' due to 
searching best parents and allowing reproduction between them. 
As for substitutions in the agents' strategy pools, in the 
schemes (a) and (c) parents give space for their offsprings 
to live and develop without the need to fight for the limited 
resources with their parents, a sacrifice for improving the survival 
of the species. Examples of parents dying after reproduction are 
numerous in nature. On the other in decision making environments 
the interpretation of killing the parent strategies is that old 
strategies - unable to lead into success - are removed to give 
way to hopefully better strategies. Schemes (b) and (d), with 
parents being saved and agents getting rid of their worst strategies, 
bear some resemblance with ``natural selection'' of the fittest surviving 
species. In decision making situations these schemes correspond 
to agents eradicating their loosing strategies. Thus it is expected 
that schemes (b) and (d) lead to tightening competition between 
agents. Furthermore, it could be expected that the scheme (d) is 
the most efficient one, because it removes the worst strategies 
and replaces them with the crossovers of the best ones, while 
saving the so far best two strategies in the game. In order to study
the effects of stiff competition between agents with continuously
improving strategies, in more detail, large scale simulations
are needed. In these simulations it turns out that when agents use  
genetic operands, the scaled utility of the system increases and 
tends to maximise with different rates depending on the mechanism 
and the parameters of the game. 

It should be noted that our mechanisms of evolution based on the 
genetic algorithms are considerably different from the mechanisms 
applied before within the framework of the minority games 
\cite{challet1,challet3,li1,li2}. Here, the strategies are changed 
by the agents themselves and they belong to the same strategy space 
$\Omega$ whose size and dimension do not change. 

\section{Results}

\subsection{Comparison between adaptation mechanisms}

\begin{figure}
\epsfig{file=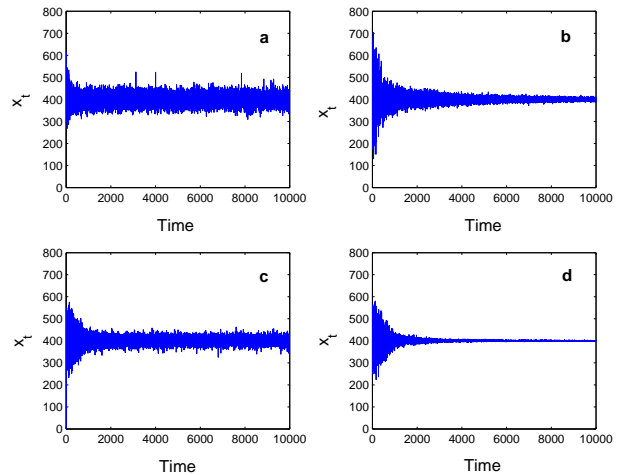,width=3.2in}
\caption{ Plots of $x_t$ (the number of agents making a particular 
action) as a function of time, for the four adaptation mechanisms 
(a)-(d), mentioned in the text.
\label{fig3}}
\end{figure}

In order to compare the above discussed four genetic adaptation 
mechanisms, we have first studied the quantity $x_t$, which describes the 
number of agents taking a particular action, of the two possible ones, 
as a function of time. The results for $N=801$ agents with $M=6$ 
memories, $S=16$ strategies, and adaptation time $\tau=40$ for 
the worst performing fraction $n=0.4$ of the agents, are depicted 
in Fig. \ref{fig3}. 
First in Fig. \ref{fig3}(a) we present the results of the adaptation
scheme (a), i.e. One Point Genetic crossover between two randomly
chosen parent strategies replaced then with the reproduced offsprings.
In this case it turned out that fluctuations in $x_t$ around its mean 
($\approx 400$) decay very rapidly from the initial level, which 
corresponds to the amount of fluctuations of the basic minority game, 
to a more or less constant level less than half of the initial level. 
This renders our scheme (a) game more efficient than the basic minority game. 
Second in Fig. \ref{fig3}(b) we present the results of adaptation scheme 
(b), i.e. One Point Genetic crossover between two randomly chosen 
parent strategies with the reproduced offsprings then replacing the 
two worst strategies of the agent's pool. In this case we observe 
that fluctuations in $x_t$ around the mean decay, first rapidly below 
the value produced by scheme (a) and then slower to very small values. 
Thus the efficiency of the system is further improved. Third in 
Fig. \ref{fig3}(c) we present the results of the adaptation 
scheme (c), i.e. Hybridized Genetic crossover mechanism between 
the best two strategies as parents, replaced after crossover 
with their offsprings. In this case fluctuations in $x_t$ around 
the mean once again decay rapidly then seemingly stabilizing to a 
level which is smaller than for adaptation scheme (a) but larger than 
for adaptation scheme (b). Fourth in Fig. \ref{fig3}(d) we present 
the results of adaptation scheme (d), i.e. Hybridized Genetic crossover 
mechanism between the best two strategies as parents and then the 
reproduced offsprings replacing the two worst strategies of the agent's 
pool. In this case we see that fluctuations in $x_t$ die off very 
rapidly, thus making the system most efficient.

\begin{figure}
\epsfig{file=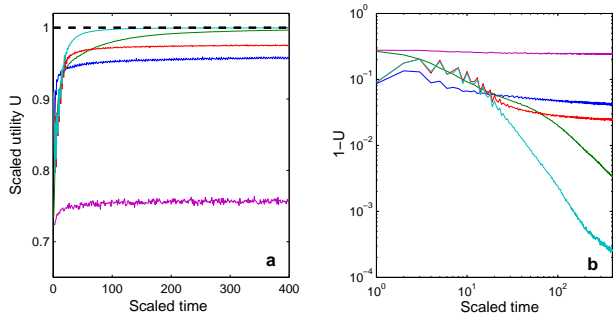,width=3.2in}
\caption{The scaled utility $U$ in panel (a) and ($1-U$) in log-log scale in 
panel (b), for the four adaptation mechanisms as functions of scaled time 
(one unit of scaled time corresponding to a time average over a bin of 50 
simulation time steps). Each curve is an ensemble average over 100 runs. 
In each panel, the magenta curve represents the basic minority game, the blue 
line represents adaptation machanism (a), green line (b), red line (c), and 
cyan line (d).
\label{fig4}}
\end{figure}

Next we focus our attention to the scaled utility $U(x_t)$, defined 
in Eq. \ref{eq1}, which is expected to give insight not only to the 
efficiency of the basic minority game and games with different adaptation 
mechanisms, but also their dynamical behaviour. Instead of the standard 
practice of studying the variation of $\sigma^2/N$ versus $2^M/N$, where 
$|\sigma|$ stands for the difference in the number of agents between the 
majority and minority groups, we study the function $U$. This is because 
fluctuations in $x_t$ decay strongly for adaptation mechanisms (b) and (d), 
in the latter case sometimes even disappearing completely. In Fig. 
\ref{fig4} we show the results of the scaled utility $U$ as a function 
of the scaled time for the four adaptation mechanisms, with the same set 
of paramaters as before ($N=801$, $M=6$, $S=16$, $\tau=40$, and $n=0.4$), 
such that panel (a) is presented in the linear scale and panel (b) $1-U$ 
in the log-log scale to see differences better. We find that the scaled 
utility rapidly saturates for the basic minority game, at a value which 
is considerably less than the maximum that can be achieved. On the other 
hand it is clearly seen that our four adaptation mechanisms greatly 
enhance the utilities close to the maximum. The value at which $U$ 
saturates depends on the mechanism of adaptation and also on the 
parameters. In Fig. \ref{fig3} it was observed that the efficiencies of 
mechanisms (b) and (d) are continually improving over time. This is 
clearly reflected in Fig. \ref{fig4}, where in the left panel the 
scaled utilities for mechanisms (b) and (d) approach asymptotically 
unity, and in the right panel the quantity $1-U$ reveals the asymtotic 
behaviour for mechanism (d) to be the fastest. Therefore, we can 
conclude that the adaptation mechanisms with parent strategies saved are 
worthwhile, and best result is achieved with the adaptation mechanism (d), 
i.e. with the elitist scheme. Later we will investigate in detail the 
parametric dependence of mechanism (d) in comparison with the simplest 
adaptation mechanism (a).  

In order to examine the evolution of strategies in the agents' pools we use 
the Hamming distance, which serves as a measure how similar the strategies 
in the pools are. The Hamming distance between two strategies is defined 
as the ratio of the uncommon bits to the total length of the strategy,
denoted by $d_H$. The strategies are said to be ``correlated'', if all the 
bits are pairwise the same, i.e. $d_H=0$; ``anticorrelated'', if all bits 
are opposite, i.e. $d_H=1$, and they are ``uncorrelated'' when exactly one 
half of the bits differ, i.e. $d_H=0.5$.

\begin{figure}
\epsfig{file=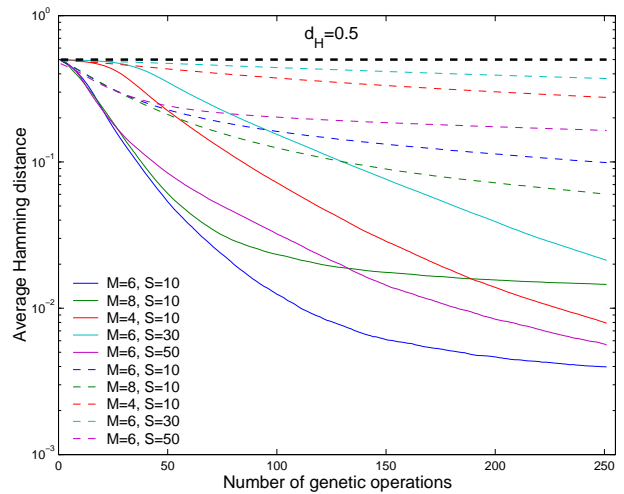,width=3.2in}
\caption{Evolution of the average Hamming distance of the agents as a 
function of the number of genetic operations (one genetic operation takes 
place after every $\tau$ time steps) for different adaptation mechanisms 
and parameters. The simulation was done with $N=801$, $\tau=80$, $n=0.4$ 
and five different sets of memories $M$ and strategies $S$. Each point 
is an ensemble average over $20$ runs. The dashed lines represent the 
results for adaptation scheme (b) and the solid line for adaptation scheme
(d). The bold dashed line is for average Hamming distance$=0.5$.
\label{fig6ja11}}
\end{figure}
Let us now consider the average Hamming distance, which for the whole
strategy space is calculated by first taking the average of
the Hamming distances over all possible strategy pairs in the agent's pool,
and then taking the average over all the agents. While it is obvious that 
individual Hamming distances between pairs of strategies can change 
as a result of genetic crossovers, the situation is more complex for the
overall average Hamming distance. As a matter of fact in the adaptation 
schemes in which the parent strategies after crossover are replaced by 
their offsprings (i.e. schemes (a) and (c)), all the bits in the agent's 
strategy pool and for that matter in the whole strategy space  
remain the same and thus no change in the average Hamming distance can 
take place. So this measure is useful only for the games where the bits 
in a strategy pool can change over time, i.e. adaptation schemes (b) and (d)
depicted in Fig. \ref{fig6ja11}. Here it is seen that as the game evolves, 
the average Hamming distance decreases in both cases towards small values, 
but for scheme (b) game this happens considerably slower than for the 
scheme (d) game. In the latter case $d_H$ reaches very small values, 
indicating that each agent tends to end up using a particular strategy 
in its pool for best performance. In the case of scheme (b) the same tendency 
exists but one would have to wait at least an oder of magnitude longer 
to achieve the same level of singularity among strategies. On the other hand 
the plots of $x_t$ (i.e. the number of agents choosing a particular action,
depicted in Fig. \ref{fig3}) shows that these strategies are such that 
the total utility, and thus the efficiency of the system tends to maximize. 
In Fig. \ref{fig6ja11} we have depicted results of varying the 
memory size $M$ and the number of strategies $S$ in each agent's pool.  
We can observe that in the case of adaptation mechanism (b) 
increasing $M$ while keeping $S=10$ fixed makes the decay in the average
Hamming distance faster, yielding $M=8$ case the fastest decaying, overall
for adaptation scheme (b). On the other hand, increasing $S$ and 
keeping $M=6$ fixed does not seem to yield systematic behaviour, 
while $S=10$ case seems to give rise to the fastest decay in the 
average Hamming distance. In the case of adaptation mechanism (d) the 
situation is even less systematic, since increasing $M$ and keeping 
$S=10$ fixed yields $M=6$ case the fastest decaying, overall for 
adaptation scheme (d), and increasing $S$ and keeping $M=6$ yields 
$S=50$ case the fastest decaying.  

\begin{figure}
\epsfig{file=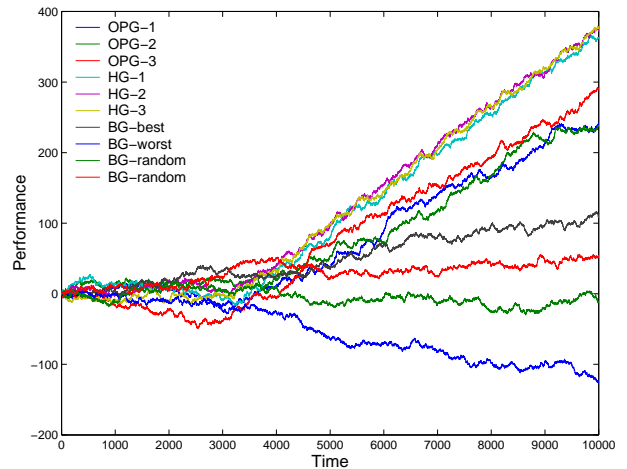,width=3.2in}
\caption{Performances of selected agents as a function
of time for a game with multiple type agents. The performances in the 
figure are scaled such that the mean performance of all the agents 
is zero. At $t=3120$, six agents begin to modify their strategies 
such that three of them used hybridized genetic crossover mechanism 
with parents saved (scheme (d)) and the other three used one point 
genetic crossover with parents are killed (scheme (a)). All the 
rest of the agents (only the performances of four agents are shown) 
played the basic minority game without adapting. The best, the worst 
and two randomly chosen agents from those who do not adapt are plotted. 
Simulations were done with $N=801$, $M=8$, $S=16$, $n=0.3$ and $\tau=80$.
\label{fig10}}
\end{figure}

Next we study a ``test'' situation to investigate whether the genetic
operations can increase the performance (i.e. the number of times an agent 
wins) of individual agents in an environment where few agents are allowed 
to modify their strategies while the others continue using the predetermined 
set of initial strategies playing the basic minority game. At the beginning 
all the agents play the basic minority game and after $t=3120$ simulation 
time steps three of the agents begin to adapt using hybridized genetic 
crossover with parents saved (adaptation scheme (d)) and another three agents 
using one point genetic crossover with parents killed mechanisms (adaptation 
scheme (a)). All the other agents continue playing the basic minority game. 
It turns out that adaptive agents, although some of them were the worst 
agent at the beginning, outperform all the other agents. Furthermore, 
we observe that all the agents who use the hybridized genetic crossover 
mechanism perform better than those using the one point genetic 
crossover mechanism. Nevertheless, the competition in these three agent 
groups is severe, as clearly seen in Fig. \ref{fig10}. The success rate,
the slope of the performance curve, is clearly different between
adaptive and other agents; the best agents in the basic game stay
far below the adaptive ones. The curves are scaled such that the
average performance of all the agents stays zero. Furthermore it
is seen in this figure that the adaptive agents stay initially
in the neighbourhood of the average performance, but crossovers
lead them quickly to success.

\subsection{Parametric studies}

Let us now move on to analyse the dependence of our adaptive games on the 
model parameters: memory size $M$, number of strategies in the pool $S$, 
crossover time $\tau$, and fraction of worst performing agents $n$. 
Here we will concentrate mainly on two different adaptation mechanisms: 
one-point genetic crossover with offsprings replacing parents (scheme (a)), 
and hybridized genetic crossover with the offsprings replacing the 
two worst strategies in the agent's pool (scheme (d)). 

\begin{figure}
\epsfig{file=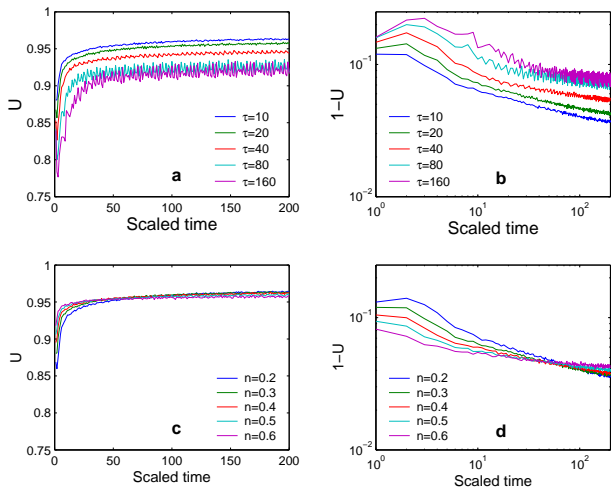,width=3.2in}
\caption{(a) Scaled utility $U$ vs. scaled time when the crossover time 
$\tau$ is varied and the fraction of the worst performing agents was 
kept at $n=0.3$, (b) the same for the quantity $1-U$ in the log-log scale, 
(c) $U$ vs. scaled time when the fraction $n$ of the worst performing 
agents is varied the crossover time kept at $\tau=10$, and (d) the same 
for the quantity $1-U$ in the log-log scale, using the adaptation
scheme (a) i.e. one-point genetic crossover with parents killed, 
and for $N=1001$, $M=5$, $S=10$. Each unit of scaled time is a time 
average of bins of $50$ time-steps and each curve is an ensemble average 
over $50$ runs. 
 \label{fig7}}
\end{figure}

In Fig. \ref{fig7} we present the results of changing the crossover time 
$\tau$ and the fraction $n$ of the worst performing agents.
The series of plots in Fig. \ref{fig7} illustrates the effect of
changing the crossover time $\tau$ and the value of the fraction $n$ of 
the worst performing agents. In this case the strategies have been 
modified using one point genetic crossover mechanism where offsprings 
replace the parents. In panel (a) of this figure we have plotted 
the total utility $U$ for five different values of $\tau$,  
while the other parameters were kept fixed, and in panel (b) 
we have plotted the same data for quantity $1-U$ in the log-log 
scale, since from this panel it is easier to compare the efficiencies 
at the end of the simulation. In panels (c) and (d), we have varied the 
fraction $n$ from $0.2$ to $0.6$ in increments of $0.1$ units, while
the rest of the parameters are kept fixed and same as in panels (a) 
and (b). We find in panel (a) that as $\tau$ increases it takes longer
time for $U$ to saturate and the efficiency decreases. On the other
panel (c), changing $n$ does not have significant effect on the
behaviour of the scaled utility.

\begin{figure}
\epsfig{file=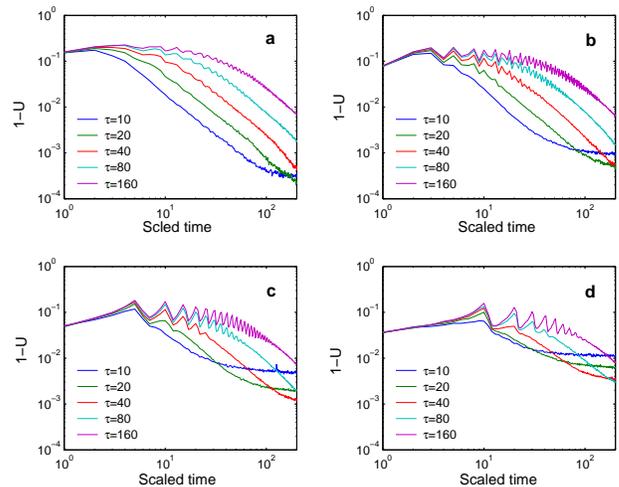,width=3.2in}
\caption{Plots of $1-U$ against scaled time in log-log scale for 
different values of $\tau$, and $M$: (a) $M=5$, (b) $M=6$, 
(c) $M=7$, (d) $M=8$, for the adaptation scheme (d), and with 
parameters $N=1001$, $S=10$, $n=0.3$. 
Each curve is an average over $50$ runs. Each unit of scaled time 
is a time average over a bin of $50$ time-steps.
\label{fig12}}
\end{figure}

\begin{figure}
\epsfig{file=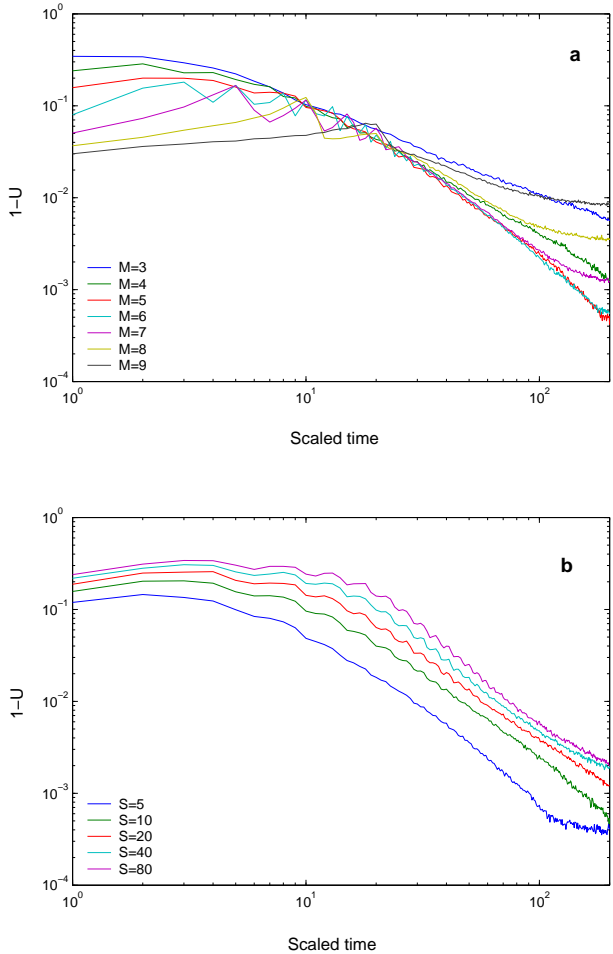,width=3.2in}
\caption{Plots of $1-U$ vs. scaled time in log-log scale for (a) 
different values of $M$, and (b) different values of $S$,for the
adaptation scheme (d), and with 
parameters $N=1001$, $M=5$, $\tau=40$, $n=0.3$. Each curve is an 
average over $50$ runs, and the unit of scaled time is a time 
average over a bin of $50$ time-steps.
\label{fig13}}
\end{figure}

In order to investigate the dependency of the utility quantity $1-U$ 
on the memory and on the crossover time, we have carried out extensive 
simulations for different $M$ and $\tau$ values, depicted in Fig. 
\ref{fig12}. In this case the memory is increased from $M=5$ in panel 
(a) to $M=8$ in panel (d). We observe that the decay rate between different adaptation 
times remain quite the same, and there exists a threshold after which 
the decay slows down. It is interesting to note that longer 
adaptation times lead to higher efficiency as the memory size increases. 
This is due to the fact that for higher dimensional strategy space it 
takes longer time until sufficiently many histories are gone through to 
verify the success of a particular strategy. On the other hand if we 
do not allow this adaptation to happen, strategies are changed too 
often and even the good ones are likely to be disregarded. 
Inspired by this observation we have studied also the effect of 
adaptation time to the final utility value which depends 
on the simulation time as we will discuss later.

In Fig. \ref{fig13} we have studied the effect of changing the memory 
size $M$ and the number of strategies $S$ on the quantity $1-U$. 
In panel (a) as $M$ increases we first observe an increase in
efficiency, but then it starts to decrease with further increase in $M$.
In panel (b) as $S$ increases the efficiency decreases. The decay rate
remains almost the same for all $S$ values as can be seen from the
slopes of the curves.

\begin{figure}
\epsfig{file=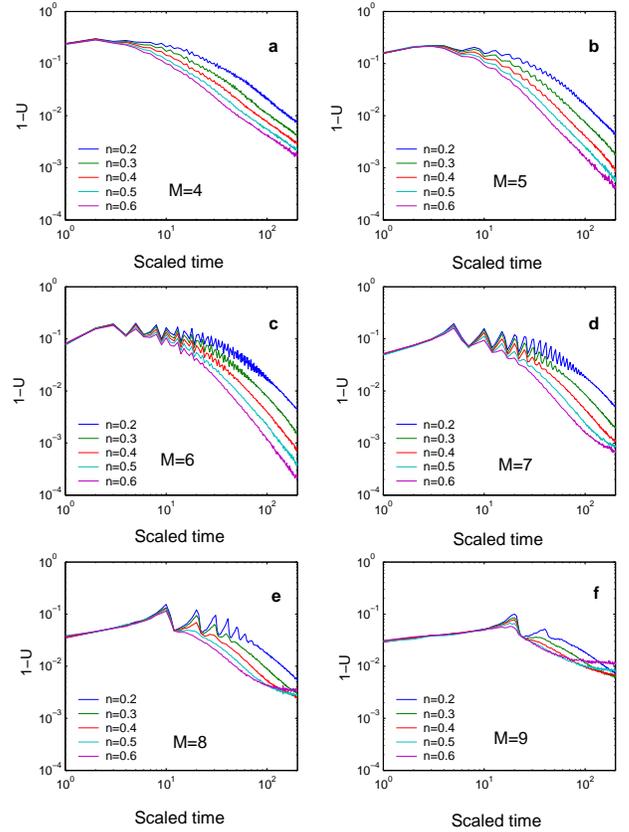,width=3.2in}
\caption{Plots of $1-U$ vs. scaled time in log-log scale for  
different $n$ and $M$, for the adaptation scheme (d).
The parameters used in the simulation are
$N=1001$, $S=10$ and $\tau=80$. Each curve is an average over $50$ runs,
and the unit of scaled time is a time average over a bin of $50$ 
time-steps.
\label{fig16}}
\end{figure}

Next we study what is the effect of changing memory to the scaled
utility as the worst fraction is varied from $n=0.2$ to $n=0.6$ and other
parameters remain fixed. Results can be seen in Fig. \ref{fig16},
from which we find that the scaled utility increases as $n$ increases
for moderate $M$ values, i.e. from $4$ in panel (a) to $6$ in panel (c). 
This behaviour changes for higher $M$ values, and shows that the system 
can reach a more efficient state, if the fraction of adaptive agents 
at any time is not too high. 

\begin{figure}
\epsfig{file=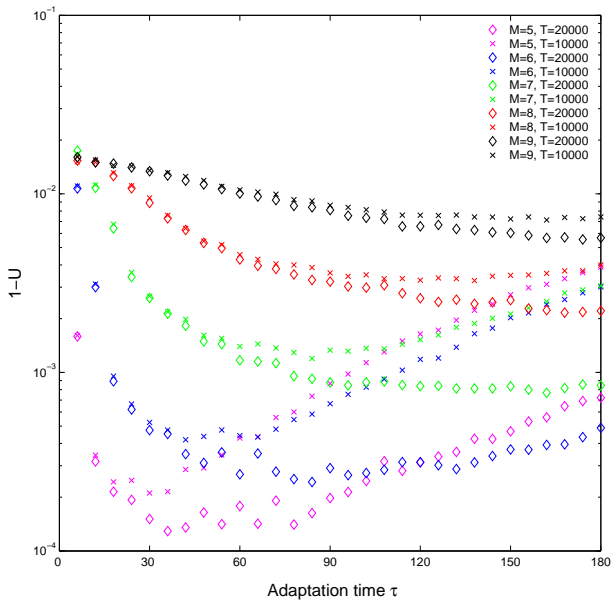,width=3.2in}
\caption{Plot of $1-U$ as a function of the adaptation time $\tau$.
Each point is calculated using the time average of the scaled utility 
from the last $500$ time-steps of the simulation for different
memories $M$ and different simulation times (indicated by $T$).
Curves are ensemble averages over 70 runs.
\label{fig14}}
\end{figure}

As stated before, we have found that the total utility, 
at the end of the game, varies with different crossover times, while
other parameters were kept fixed. Especially, larger memory values
increase the dimension of the strategy space and require longer 
adaptation times. In order to study this dependency in more detail 
we have simulated the hybridized genetic crossover with parents saved 
mechanism (i.e. scheme (d)) for several $M$ values and different 
values of total simulation time $T$. Results are illustrated in Fig.  
\ref{fig14}. The minima in the curves are found dependent on the 
simulation time, as can be seen by the slight differences between 
the curves for same value of $M$ but different values 
of $T$. If the simulation times were unrestricted 
there would not exist an increase in the curves as adaptation 
time increases. The longer the agent can observe its strategies 
the more certain it can be of their mutual performance. If the 
adaptation time is reduced too much, crossovers take place more at 
random. Thus it could be envisaged these curves to give guidance 
for a preferable adaptation time. Intuitively, one could guess that 
a good adaptation time would be close to $2^M$, because if the 
occurrence of a history were uniformly distributed, this would constitute
the expectation time for an agent to go through all the histories 
once and thus see how successful a response determined by a strategy 
has been in each case. 

\section{Discussions}

In this paper we have applied various adaptation mechanisms based on 
genetic algorithms within the framework of the minority game
and found significant changes in the collective and individual
behaviour of the agents. We found that the hybridized genetic
crossover mechanism in which the best strategies are chosen as parents
and their offsprings replace the two worst strategies in the agent's pool
leads the system fast towards a state where the scaled utility tends to 
its maximum. This mechanisms is clearly better than those where the
parent strategies are chosen randomly and those where parents are 
replaced by their offsprings. The pre-eminence of the hybridized 
mechanism can be seen on the system as well as on the agent level: 
fluctuations in $x_t$ smooth down quickly and the 
agents outperform those using other mechanisms participating the same 
game. The success of genetic algorithm based adaptation mechanisms 
in minority games is interesting and suggests its use also in other 
game theoretic optimization problems. It should be noted, that 
the minority game deviates from the traditional optimization problems 
because it does not include a particular object function or functions 
that are tried to be maximized. This makes our finding even more 
interesting, exposing a certain characteristic of the minority game: 
if agents have the possibility to adapt trough the responses to 
the stimuli, they drive towards a state where their own performance 
improves and the collective of all participants gain maximum amount 
of utility every time the game is played. This property is not trivial 
to understand but it bases on the convergence of strategies in the 
strategy space towards, in a way, the optimal ones. They are optimal 
just in the sense that they tend to bring the maximum utility for 
the collective, meaning that at each time step the number of agents 
who win is as large as possible: the number of satisfied individuals 
is at maximum. Adaptation mechanism further extends the class of 
phenomena, minority games are roughly able to describe. This is because 
the basic minority game lacks an efficient learning mechanism, but 
still the number of systems where individuals try to improve their 
performances in competitive environments is huge. Our way to include 
adaptation is not arbitrary but has analogies 
with reality and is based on learning and
combination of different adaptation schemes, a way that is very simple and
common in nature and social systems.

\begin{acknowledgments}
This research was partially supported by the Academy of
Finland, Research Centre for Computational Science and Engineering,
project no. 44897 (Finnish Centre of Excellence Programme 2000-2005).
\end{acknowledgments}

\end{document}